\documentclass{article}
\usepackage{cite}
\usepackage{amssymb}
\usepackage{times,bbm}
\usepackage{graphics,dcolumn,bm,float}
\usepackage{amssymb,amsmath,rotate,color}
\usepackage{mathtools}
\usepackage{booktabs}
\usepackage{rotating}
\usepackage{float}
\usepackage{braket}
\usepackage{amsmath}

\usepackage{graphicx}

\newcommand{\bmt}{\left[\begin{matrix}}
\newcommand{\emt}{\end{matrix}\right]}

\newcommand{\ee}{\end{equation}}

\newcommand{\eea}{\end{eqnarray}}

\newcommand{\spinup}{\psi_{\scriptscriptstyle\uparrow}}

\newcommand{\spinups}[1]{\psi^#1_{\scriptscriptstyle\uparrow}}
\newcommand{\spindown}{\psi_{\scriptscriptstyle\downarrow}}

\newcommand{\spindowns}[1]{\psi^#1_{\scriptscriptstyle\downarrow}}
\newcommand{\bigt}{\bigtriangleup}

\newcommand{\bfrp}[1]{{({\bf r}^\prime )}}

\newcommand{\appsection}[1]{\let\oldthesection\thesection
  \renewcommand{\thesection}{Appendix \oldthesection}
  \section{#1}\let\thesection\oldthesection}

\newcommand{ \sech}{\rm sech}
\begin{document}
\def\twidle{\widetilde}
\def\f{\frac}
\def\omit#1{_{\!\rlap{$\scriptscriptstyle \backslash$}
{\scriptscriptstyle #1}}}
\def\vec#1{\mathchoice
    {\mbox{\boldmath $#1$}}
    {\mbox{\boldmath $#1$}}
    {\mbox{\boldmath $\scriptstyle #1$}}
    {\mbox{\boldmath $\scriptscriptstyle #1$}} }
\def\eqn#1{Eq.\ (\ref{#1})}
\def\boxit#1{\vcenter{\hrule\hbox{\vrule\kern8pt
      \vbox{\kern8pt#1\kern8pt}\kern8pt\vrule}\hrule}}
\def\Boxed#1{\boxit{\hbox{$\displaystyle{#1}$}}} 
\def\sqr#1#2{{\vcenter{\vbox{\hrule height.#2pt
        \hbox{\vrule width.#2pt height#1pt \kern#1pt
          \vrule width.#2pt}
        \hrule height.#2pt}}}}
\def\square{\mathchoice\sqr34\sqr34\sqr{2.1}3\sqr{1.5}3}
\def\Square{\mathchoice\sqr67\sqr67\sqr{5.1}3\sqr{1.5}3}
\def\lambdabar{{\mathchar'26\mkern-9mu\lambda}}
\def\thrdotovervx{\buildrel\textstyle...\over v_x}
\def\thrdotovervy{\buildrel\textstyle...\over v_y}
\title{ Current Density of Majorana Bound States}
\author{{\small \ Mehran Zahiri Abyaneh}\footnote{me\_zahiri@sbu.ac.ir} \ {\small
        and Mehrdad Farhoudi}\footnote{m-farhoudi@sbu.ac.ir} \\
        {\small Department of Physics, Shahid Beheshti University,
        Evin, Tehran 19839, Iran}}
\date{\small September 24, 2022}
\maketitle
\begin{abstract}
\noindent
 It is known that a non-local complex fermion can be written in
terms of two Majorana fermions. We exploit this fact to explain
the system of two Majorana zero modes bound to a vortex and an
anti-vortex, on the surface of a topological insulator in contact
with an s-wave superconductor, as a non-local complex fermion.
Although the current density of a single zero mode vanishes, by
starting with a wave packet consisted of the positive and negative
energy complex fermions, we specify that a time-dependent
oscillatory motion emerges in the system. We also show that the
amplitude and frequency of the oscillations depend on the relative
distance of those two zero modes. Therefore, the observation of
this oscillatory motion can be considered as a signature of the
Majorana zero modes. Also, as the frequency of such an oscillatory
motion depends on the distance between the two zero modes, it can
be adjusted to bring this frequency within the resolution of
observations. Furthermore, we indicate that the predicted
oscillatory current is the reminiscent of the
\emph{zitterbewegung} effect.

\end{abstract}
\bigskip
\section{Introduction}
\indent

Unlike the standard solutions of the Dirac equation, i.e. the
electron and positron, a solution of the Majorana equation is its
own antiparticle. While such a particle has~not been detected in
particle physics, in solid states physics a pair of localized
Majorana zero-energy modes (MZMs) have been predicted to reside at
the core of vortices in the superconductor-topological insulator
(STI) system~\cite{Jackiw1,FuKane,Jackiw2}. Among different setups
to realize MZMs, one can mention a p-wave superconductor with a
non-relativistic kinetic term and a vortex order
parameter~\cite{Read-Green2000}. From theoretical point of view,
it is predictable that MZMs will have very important role as
qubits in the topological quantum computing in
future~\cite{Ivanov}. The susceptibility of qubit systems to
decoherence, as the principal obstacle in realizing a scalable
quantum computer, has led some efforts to detect MZMs emerging in
topologically non-trivial superconducting phases in order to build
fault-tolerance quantum computing. Indeed, in such a quantum
computer, information is stored in a system of two MZMs located
faraway from each other and thus, would be more protected against
local perturbations that may cause quantum
decoherence~\cite{Kitaev,Nayak,Stern,Lahtinen}. Also, the
adiabatic braiding of MZMs can be used to perform qubit
operations, while their fusions provide the means of qubit read
out~\cite{Yazdani}. The enthusiasm about MZMs has led to many
theoretical works on this subject in the literature, see, e.g.
Refs.~\cite{Yazdani,Gurarie-Radzihovsky,Tewari-etal,Bergman-LeHur,Ghaemi-Wilczek,Forlov}.

On the other hand, it is known that Majorana's original
work~\cite{Majorana1937} was~not restricted only to MZMs, which
are static solutions of the Majorana equation. His approach was
quite general and he introduced whole field covering the entire
energy-range, called Majorana fermions (MFs), which are solutions
of the Dirac equation in a specific representation. MFs are their
own anti-particles and include the entire energy-range, and are
the topic of research in particle physics context. One of the main
candidate of being a MF is
neutrino~\cite{Abbas-Abyaneh1,thesisAbyaneh,Abbas-Abyaneh2,Abdousallam-Abyaneh},
and in rare nuclear decays, several highly sensitive experiments
around the world are going on to search for any evidence to
illustrate that neutrinos are
MFs~\cite{Alessandria2011rc,Auger2012ar}. If neutrinos are MFs,
the dilemma of smallness of their masses can then be explained via
the see-saw
mechanism~\cite{GellMann:1980vs,Yanagida:1979as,Glashow:1979nm,Mohapatra:1979ia},
which predicts that there are two kinds of neutrinos. The light
left-handed neutrinos and the heavy right-handed ones, where the
higher the mass of the right-handed neutrinos, the lower the mass
of the left-handed ones. In this respect, recently it has also
been claimed that neutrinos can be regarded as the Bogoliubov
quasi-particles (Bqp)s~\cite{Fujikawaa}. Nevertheless, Majorana
neutrinos have been elusive so far in the particle physics
experiments\rlap.\footnote{The mass of right-handed neutrinos can
be as high as $10^{15}\, GeV $ in the see-saw
scenario~\cite{thesisAbyaneh}.}

Despite the fact that MZMs are much studied on the theoretical
side and the scientific excitement around those is on a par with
gravitational waves and the Higgs boson, the issue of their
detection has some difficulties. Several groups have reported the
detection of Majorana bound states (MBSs) in nanowires through a
measurement of a zero bias peak in tunneling spectroscopy
experiments, see, e.g., Ref.~\cite{Mourik}. However, the research
community is still skeptical towards experimental findings in this
regard~\cite{Yazdani,Forlov}. The fact that MFs are their own
anti-particles and charge-less makes those elusive for unambiguous
detection in experiments, and subtle schemes are needed for
indirect but conclusive signatures of their presence. In this
vein, experiments are redesigned to probe other MZMs properties
such as their particle-hole symmetry and spin, which might lead to
clear results~\cite{Peng,Feldman,Jeon,Li}. There are also many
proposals to detect MFs based on interferometric structures, for
example using a two-terminal Mach-Zehnder
setup~\cite{Kane2009,Akhmerov}, or via measuring the energy of the
bound state~\cite{Jackiw2}. Moreover, neither the braiding nor the
fusion of vortices has been realized in the laboratory, and a
variety of theoretical plans have been proposed to demonstrate the
appearance of non-Abelian anyons in a topological
superconductor~\cite{Beenakker2019}.

To address these advances, we investigate the current density of
MBSs in the STI system. To perform such a task, we exploit the
fact that one cannot talk about the state of a single MZM since it
contains only half a fermion. The only
physical observables are the fermionic occupation numbers, which
consist of two MZMs~\cite{Alicia,Lijense}. By utilizing a
fermionic field made of two MBSs bound to a vortex and an
anti-vortex, we calculate the current density of the field and
specify that the interference of the positive and negative energy
parts of it leads to a time-dependent oscillatory motion in the
system. We also show that the amplitude and frequency of the
oscillations depend on the distance between those MZMs.

This issue is somehow similar to the \emph{zitterbewegung} (ZBW)
-- a trembling/quivering motion -- which was introduced by
Schr\"odinger~\cite{schrodinger30,schrodinger31,barut-etal81a}.
Such a phenomenon has been studied, for instance, for a
wave-packet made of positive and negative energy
electrons~\cite{Huang}, an electron in the presence of an external
magnetic field in commutative space~\cite{zahirifarhoudi} and
separately for the non-commutative
phase-space~\cite{zahirifarhoudi1,zahirifarhoudi2} and references
therein. In this respect, in Ref.~\cite{Huang}, it has been shown
that the current produced by the ZBW actually causes the intrinsic
magnetic moment of a Dirac fermion particle and hence, the total
magnetic moment of electron is produced by both the orbital and
the intrinsic angular momenta with the correct gyromagnetic g
factor. Thus, theoretical understanding of the ZBW may shed light
on the nature of spin of elementary particles, see, e.g.,
Refs.~\cite{BarutZanghi84,Pavsic1993,rusin-etal2006,hestenes2009,Awobode}.
However, despite increasing evidence that ZBW is real and, in
principle, observable (e.g., in a Bose-Einstein
condensate~\cite{heszitter2}, in crystalline
solids~\cite{Zawadzki}, semiconductors~\cite{Zawadzki2011},
graphene~\cite{Rusin2013} and in silicene~\cite{Romera} ), it has
always been challenged and become an unpleasant
aspect~\cite{Deriglazov}, since Foldy and Wouthuysen (FW) observed
that, in absence of external fields, the ZBW can be avoided via
the transformation employed by them~\cite{Foldy}. Indeed, the FW
transformation is a unitary transformation that transforms the
Dirac Hamiltonian for a localized free particle into a
(block-diagonalized) Hamiltonian in which positive and negative
electron energies are decoupled, see, e.g.,
Refs.~\cite{Costella,Strange}. In this regard, as the ZBW goes
hand in hand with the existence of negative and positive energy
solutions\rlap,\footnote{Alternatively, in
Refs.~\cite{hestenes90,Kobakhidze}, it has been stated that the
ZBW provides a physical interpretation for the complex phase
factor in the Dirac wave function. Also, see
Ref.~\cite{David2010}, wherein it has been shown that the ZBW
cannot be described by only one frequency.}\
 it has been claimed to be a
frame-dependent concept, and hence not observable. Nevertheless,
there have been plenty of works devoted to this issue even since
then, see, e.g.,
Refs.~\cite{zahirifarhoudi,Eckstein2017,Hestenes2018,Reck2020,Silenko,Kobe}
and references therein. However, in the case of electrons and
positrons, the ultrahigh frequency of ZBW is $f_{_{\rm ZBW}}\! =
2mc^2/h \sim 10^{20}\, Hz$ and amplitude given by the Compton
wavelength $\lambda_c \sim 10^{-13}\, m$, whose direct measurement
is still beyond experimental capabilities, and indeed, the lack of
empirical evidence is due to the transient nature of wave-packet
ZBW.

The outline of the work is as follows. In the next section, we
briefly review the Dirac equation, the ZBW phenomenon and the FW
transformation. Sec.~3 is devoted to the physics governing the
surface of topological insulator in the proximity of an s-wave
superconductor, wherein its corresponding Hamiltonian is studied.
We indicate that in this system the ZBW of the relevant surface
excitations can avoid the FW objection. Furthermore, we study the
symmetries of the Hamiltonian and show that only the gauge
symmetry is conserved when the superconducting pairing is present
in the system. In Sec.~4, we right down the zero-energy solutions
of the STI system Hamiltonian and indicate that a wave-packet
consisted of two MBSs accommodates a non-vanishing current density
whose spatial components represent the ZBW while its frequency can
be adjusted for detection. Finally, we conclude the summary of the
results in Conclusions.

\section{ Dirac Equation and Foldy-Wouthuysen Transformation }\label{FW}
\indent

The Dirac Lagrangian for a free electron with mass $m$ is written
as
\begin{eqnarray}\label{24}
{\cal L}=\bar{\Psi}(i\gamma^\mu\partial_\mu -m){\Psi} ,
\end{eqnarray}
where we have used the natural units $\hbar=1=c$, $\bar{\Psi}=\Psi^{\dagger}\gamma_0$
and $\gamma_{\mu}$s are the Dirac gamma matrices that, in the Weyl
or chiral representations, are defined as
\begin{equation}\label{D}
\gamma_i=\left(
                 \begin{array}{cc}
                   0 & \sigma_i \\
                 - \sigma_i & 0 \\
                 \end{array}
               \right) ,\qquad \gamma_5=\left(
                 \begin{array}{cc}
                   \mathbb{I}_2 & 0\\
                 0 &- \mathbb{I}_2 \\
                 \end{array}
               \right)\,
               \qquad\ \textrm{and}\qquad\ \beta\equiv\gamma_0=\left(
                 \begin{array}{cc}
                     0& \mathbb{I}_2\\
                   \mathbb{I}_2 & 0 \\
                 \end{array}
               \right)
\end{equation}
with $\sigma_i$s and $\mathbb{I}_2$ as the $2\times2$ Pauli
matrices and the unit matrix, respectively. In addition in terms
of the Dirac (gamma) matrices, the $\gamma_5$ matrix is
constructed as $\gamma^5=i\gamma^0\gamma^1\gamma^2\gamma^3$, and
the matrix $\mbox{\boldmath$\alpha$}$ as
$\alpha_i\equiv\gamma^0\gamma_i$ for $i=1, 2, 3$.
 Moreover, the $\gamma^{\mu}$ matrices
satisfy
\begin{eqnarray}
\{\gamma^{\mu},\gamma^{\nu}\}\equiv \gamma^{\mu} \gamma^{\nu}+\gamma^{\nu} \gamma^{\mu}=2\,\eta^{\mu\nu}, \nonumber
\end{eqnarray}
where $\eta^{\mu\nu}$ $(\mu,\nu= 0,\cdots,3)$ is the Minkowski metric in $(1+3)$
dimensions with the signature $-2$.
Now, the Dirac
Hamiltonian is
\begin{equation}\label{Diracrep}
{\displaystyle  {H}\equiv {\boldsymbol {\alpha }}\cdot \mathbf {p}
+\beta m}
\end{equation}
and the corresponding equation of
motion is the celebrated Dirac
equation~\cite{dirac2-1928}
\begin{equation}
i  \frac{\partial}{\partial t}\Psi=
(\gamma_0{\mbox{\boldmath$\gamma$}}\cdot{\bf p}+\beta m )\Psi .
\end{equation}

By defining the right-handed and left-handed projection operators
$\mathcal P_+=(1+\gamma_5)/2$ and $\mathcal
P_-=(1-\gamma_5)/2$ that project $\Psi$ to its right- and
left-handed components respectively and assuming $\Psi=\left(\!\!
                 \begin{array}{cc}
                   \psi_R \\
                 \psi_L \\
                 \end{array}
               \!\!\right)$,
the Dirac equation becomes
\begin{eqnarray}
          {\boldsymbol\sigma} \cdot \mathbf {p} \, \psi_R \!\!\! &+& \!\!\! m \,  \psi_L=  \varepsilon \psi_R  \\
  -{\boldsymbol\sigma} \cdot \mathbf {p} \,  \psi_L\ \!\!\!   &+& \!\!\! m\, \psi_R  =  \varepsilon \psi_L
\end{eqnarray}
with the eigenvalues $ \varepsilon=\pm\sqrt{{\rm p}^2+m^2 }$.
The formal similarity of this equation with the one attained from
the Bardeen-Cooper-Schrieffer (BCS) Hamiltonian of
superconductivity~\cite{BCS}, led Nambu to transport the BCS
theory to the physics of strong interactions~\cite{NJL1,NJL2}. For
a Dirac particle, ZBW arises when one computes the time-dependence
of the position operator in the Heisenberg picture, namely
\[ {\frac {\partial x_{k}(t)}{\partial t}}=i\left[H,x_{k}\right]= \alpha
_{i},\]
 where $x_k(t)$ is the position operator at time $t$.
However to obtain the value of the ZBW velocity, one needs to
calculate the expectation value $<\!\alpha\!>$ while using a
wave-packet consisted of both positive and negative energy
solutions of the Dirac equation, see, e.g.,
Refs.~\cite{zahirifarhoudi,zahirifarhoudi1,zahirifarhoudi2} and
references therein.

Nevertheless, FW first introduced the $4\times4$ unitary
transformation operator
\begin{equation}\label{FWoperator}
U=\mathbb{I}_4\,\cos \vartheta +\beta {\boldsymbol {\alpha
}}\cdot{\hat {\mathbf {p}} }\sin \vartheta ,
\end{equation}
where ${\hat{\mathbf {p}}}$ is the unit vector in momentum space
and $\vartheta$ is an arbitrary angle. Then, they acted it on a
fermion eigenket,
 \[{\displaystyle \Psi \to \Psi '=U\Psi },\]
and simultaneously on the free-fermion Dirac Hamiltonian operator
in the Dirac-Pauli representation~(\ref{Diracrep}) in the
bi-unitary fashion as
\[{\displaystyle {\begin{aligned} H\to H'\equiv U {H}U^{-1}=U({\boldsymbol
{\alpha }}\cdot \mathbf {p} +\beta m)U^{-1}. \end{aligned}}}\]
Using the commutativity properties of the Dirac matrices, this new
Hamiltonian reads
\begin{equation}\label{simplified}
{\displaystyle {\begin{aligned}{H}'&=({\boldsymbol {\alpha }}\cdot
\mathbf {p} +\beta m)(\cos 2\vartheta -\beta {\boldsymbol {\alpha
}}\cdot {\hat{\mathbf {p}} }\sin 2\vartheta ),\end{aligned}}}
\end{equation}
and hence, one gets
\[{\displaystyle {H}'={\boldsymbol {\alpha }}\cdot \mathbf {p}
\left(\cos 2\vartheta -{\frac {m}{|\mathbf {p} |}}\sin 2\vartheta
\right)+\beta (m\cos 2\vartheta +|\mathbf {p} |\sin 2\vartheta
)}.\]
 However, by choosing
\[{\displaystyle \tan 2\vartheta \equiv {\frac {|\mathbf {p} |}{m}}},\]
 this Hamiltonian reduces to
\[{\displaystyle  {H}'=\beta {\sqrt {m^{2}+|\mathbf {p} |^{2}}}},   \]
which is the Dirac Hamiltonian in the
Newton-Wigner~\cite{Strange,NewtonWigner} representation. Now, the
commutator $[x_i(t), {H}']$ is equal to the group velocity $v_g$,
and accordingly, it has been claimed that the ZBW motion is a
representation-dependent concept, which vanishes in the
Newton-Wigner representation.

However, it should be noted that the FW transformation is based on
the Newton-Wigner work~\cite{NewtonWigner}, which argues that a
state, localized at a certain point, after a translation becomes
orthogonal to all un-displaced states localized at that point.
Whereas we will show that the main peculiarity of the present work
is that the complex fermionic state is constructed out of two
separated MBSs. On the other hand, it has long been argued that
MFs are non-local (correlations incompatible with a local hidden
variable theory) in nature~\cite{Wang,Campbell,Khaymovich}.
Therefore, the Newton-Wigner argument and the FW transformation
would~not be relevant to this case, and one can expect that the
ZBW motion may~not vanish there.

\section{Topological Insulator Superconductor System}\label{formulation.sec}
\indent

The superconductivity can be induced into the surface of a
topological insulator in proximity of an s-wave superconductor.
The Hamiltonian density of such a two dimensional system is
written as~\cite{FuKane}
\begin{equation}
h=\spinups{\ast} p_-\, \spindown + \spindowns{\ast}\, p_+\,
\spinup - \mu (\spinups{\ast}\, \spinup + \spindowns{\ast}\,
\spindown) + \bigtriangleup \spinups{\ast} \spindowns{\ast} +
\bigtriangleup^\ast \spindown \spinup \label{eq:1}
\end{equation}
where $\spinups{\ast}$ denotes complex conjugate of $\spinup$, $
\mu$ is the chemical potential, $p_\pm \equiv p_x \pm ip_y$. Also,
the order parameter is a scalar as
\begin{eqnarray}
 \bigtriangleup ({\bf r}) = v(r)\, e^{i \phi},
\end{eqnarray}
where $v (r)$ is a real scalar function of the
distance and $\phi$ is the polar angle.

This Hamiltonian can be written in two-component matrix notation
as
\begin{eqnarray}
i\,\partial_t \psi = \left({\boldsymbol\sigma}\cdot{\bf p} -\mu \right) \psi
+\bigtriangleup \,i\sigma^2\;\psi^{\ast}
\label{eq:2-ins}
\end{eqnarray}
with
\begin{equation}\label{2solution}
\psi=\left(\!\!
      \begin{array}{c}
      \psi_{\scriptscriptstyle \uparrow} \\
      \psi_{\scriptscriptstyle \downarrow} \\
      \end{array} \!\!\right),
\end{equation}
 where $\boldsymbol\sigma$ represents
the two Pauli matrices $(\sigma^1,\sigma^2)$. Although a static
solution of Eq. (\ref{eq:2-ins}) can easily be found, to study the
time evolution of eigenstates, one needs to use two copies
of~(\ref{eq:1}) to get the Hamiltonian density as (in units $
v_F=1$, where $v_F$ is the Femri velocity)
\begin{equation}
{\mathcal H}=\frac{1}{2}\Psi^{\dagger} \left(\begin{array}{ccc} {\boldsymbol\sigma} \cdot \mathbf {p} - \mu & & \bigt\\[.5ex]
\bigt^\ast & & -{\boldsymbol\sigma} \cdot \mathbf {p} + \mu
\end{array}
\right)\Psi=\frac{1}{2}\Psi^{\dagger} H\Psi , \label{eq:7}
\end{equation}
whose eigenvalues are
\begin{equation}
E = \pm \sqrt{({\rm p} \pm \mu)^2 +  \bigt^2} \label{eq:15}
\end{equation}
and its eigenstates are solutions of the Schr\"odinger equation
\begin{equation}\label{schrodinger}
 H\Psi=E\Psi .
\end{equation}
The Dirac gamma matrices~(\ref{D}) can also be used to write the
Hamiltonian of the system as
\begin{eqnarray}\label{Ham} H= \alpha^j p_j+\beta \Delta, \label{2.11}
\end{eqnarray}
where $j=1,2$ and the matrix $ \Delta$ is defined as $\Delta\equiv
v(r)\, e^{i\gamma_5\phi}$.

In the superconductivity context, the Hamiltonian  $ {\mathcal H}$
is known as the Bogoliubov-de Gennes Hamiltonian, and is similar
to the Dirac Hamiltonian in particle physics
applications~\cite{FuKane}. An important point is that, due to the
existence of $\gamma_5$ in Hamiltonian~(\ref{Ham}), the FW
procedure is not applicable in the usual manner. However, it can
be brought to the form which is block-diagonal with respect to
negative and positive energy solutions by first performing a
simple canonical transformation to remove the `odd' parts of the
mass term and then applying the power series FW transformation to
the resulting Hamiltonian~\cite{Costella, Eriksen}. Once again, we
emphasis that the FW transformation would not be relevant to the
present work due to the non-local nature of the system under
study.

Solutions of Eq.~(\ref{schrodinger}) have the general form
\begin{equation}\label{majosol}
\Psi=\left(\!\!
            \begin{array}{c}
             \psi\\
            \psi^c\\
            \end{array}
          \!\!\right),
\end{equation}
where $\psi^c=\left(\!\!
            \begin{array}{c}
            \psi_{\downarrow}^*\\
          - \psi_{\uparrow}^*\\
            \end{array}
          \!\!\right)$.
 In the case $E=0$, the solution is a MZM, which satisfies the
pseudo-reality constraint
\begin{equation}\label{reality}
\mathcal{C}\Psi^\ast= \Psi ,
\end{equation}
where the charge-conjugation operator is $\mathcal{C}=i \gamma_2
K$ with $K$ as the complex-conjugation operator. By operating the
chiral projection operators ${\mathcal P}_\pm$ on $\Psi $, we
obtain
\begin{equation}\label{23}
\psi=\mathcal P_+\Psi \qquad {\rm and} \qquad \psi^c=\mathcal
P_-\Psi ,
\end{equation}
which indicates that the $\gamma_5$ also relates eigenstates with
positive energies to those with negative energies.

However when $E\neq 0$, $ \Psi$ does~not satisfies the
pseudo-reality constraint~(\ref{reality}), and general solutions
of Eq.~(\ref{schrodinger}) are known as Bqps and have the form of
Dirac four-spinors, which correspond to spin 1/2 fermions. One may
exploit these plane wave solutions,
 for $\mu=0 $,  to build the wave-packet
 \begin{equation}\label{eq:12}
\Phi_{\rm Bqp}({\bf r},t)  =  \int \frac{d^2 p}{(2 \pi)^2}
\sum_{s= 1}^{2} \left[a_{s}(p)\Phi_+^{s}({\bf p}) e^{-i(E t- {\bf
p}\cdot {\bf r}) }+ b_{s}^{*}(p)\Phi_-^{s}({\bf p}) e^{i(E t- {\bf
p}\cdot {\bf r})} \right],
\end{equation}
where $p=(E, {\bf p})$, the sign $+(-)$ stands for positive (negative)
energy, and $a_s$ and $b_s^*$ are arbitrary coefficients. Also,
the spinors $\Phi^s_{\pm}$ are given by
\begin{eqnarray}\label{solutions2}
\!\!\!\!\!\!\!\!\!\!\!\!\!\!\!\!\Phi_{+}^{s}\!\!\!\!\!\!&=&\!\!\!\!\!\frac{1}{
2\sqrt{ \bigt(E+ \bigt)}}\! \left(\!\!
\begin{array}{c}e^{-i
\phi}\left[{ E+ \bigt-\sigma^i p_i}
\right]\chi_s\\
 e^{i  \phi}\left[{  E+ \bigt+\sigma^i p_i} \right]
\chi_s\\
 \end{array}\!\!\!\right)\quad {\rm and}\qquad
\Phi_{-}^{s}\!=\!\frac{1}{2 \sqrt{ \bigt(E+ \bigt)}}\!
\left(\!\!
\begin{array}{c}-e^{i  \phi}\left[{  E+ \bigt+\sigma^i
p_i}
\right] \chi_s\\
 e^{-i  \phi}\left[{  E+ \bigt-\sigma^i p_i}
\right]\chi_s\\
 \end{array}\!\!\!\right)\!,
\end{eqnarray}
where
 \[ \chi_1 =\frac{1}{\sqrt{2}} \left(
\begin{array}{c} 1\\ e^{i \varphi} \end{array} \right)\qquad\textit{\rm and}
\qquad\chi_2 = \frac{1}{\sqrt{2}} \left( \begin{array}{c} 1\\
e^{-i\varphi}
\end{array} \right)\]
 refer to eigenvectors of ${\mbox{\boldmath$\sigma$}}
\cdot\hat{{\bf p}}$ while $\hat{{\bf p}}
=(\cos{\varphi},\sin{\varphi})$. As the spin of Bqps in
topological insulator is in-plane and is perpendicularly locked to
the momentum~\cite{Tian}, it is non-chiral. Besides, the negative
energy spinors are given by $C \Phi^\ast_{1,2}\ (-{\bf p})$, and
the normalization condition
 leads to the constraint
\begin{equation}
 \int \frac{d^2 p}{(2 \pi)^2} ~\sum_{s= 1}^{2}
\Big[\vert a_{s}(p) \vert^2 +\vert b_{s}(p) \vert^2 \Big] = 1.
\end{equation}

In the case $ \Delta=0$, with arbitrary constants, say $\eta$ and
$\zeta$, Hamiltonian~(\ref{Ham}) is invariant under
transformations
\begin{equation}\label{transform1}
\Psi\rightarrow\exp( i \eta)\Psi ,\qquad\qquad\
\bar{\Psi}\rightarrow \exp( i \eta)\bar{\Psi}
\end{equation}
and
\begin{equation}\label{transform2}
 \Psi\rightarrow\exp\big(i \gamma_5\zeta
\big)\Psi, \qquad\quad \bar{\Psi}\rightarrow\bar{\Psi}
\exp\big(i\gamma_5 \zeta \big),
\end{equation}
which lead to the conserved currents  $j_{\mu}=
\bar\Psi\gamma_{\mu}\Psi$ and $j_{\mu5}=
\bar\Psi\gamma_{\mu}\gamma_5\Psi$. The first current is the vector
current related to the gauge symmetry and the second one is the
chiral current related to the $\gamma_5$ symmetry. The relevant
continuity equations are
\begin{eqnarray}\label{25}
\partial_{\mu}j_{\mu}\!\!\! &=&\!\!\! 0 , \\
\partial_{\mu}j_{\mu5 }\!\!\! &=&\!\!\! 0 .
\end{eqnarray}
However, when $ \Delta\neq 0$, the vector current remains
conserved\rlap,\footnote{It is in order to mention that the
pairing term in Hamiltonian~(\ref{Ham}) breaks the gauge $U(1)$
symmetry and the charge conservation, but, as this symmetry
breaking happens spontaneously, it can be restored by the
Nambu-Goldstone boson~\cite{Nambu1,Nambu2,Goldstone}.}\
  although it vanishes for
MZMs\footnote{This issue is~not necessarily for the case of a MBS
consisted of a vortex and an anti-vortex as will be shown
below.}\
 \cite{Jackiw2}, whereas the chiral current is~not
conserved\cite{Jackiw2} because the Hamiltonian of the system
does~not commute with the chiral operator $\gamma_5$ due to the
$\gamma_0$ term in $ H$.

\section{ Current Density of Majorana Bound States}
\indent

In this section, we investigate the current density, $j_{\mu}=
\bar\Psi\gamma_{\mu}\Psi$, of MBSs in the STI
system.
 In this respect, we study a superconductor attached to a
topological insulator in the vortex/anti-vortex background. The
zero-energy mode for the vortex at the origin approximately
becomes~\cite{Jackiw2}
\begin{subequations}
\begin{equation}
\psi^v_0 \approx N\, e^{-i\pi/4}\ e^{-V( {\bf r})}
\left(
\begin{array}{c}
1\\
0
\end{array}
\right)
\label{eq:27a}
\end{equation}
where $N$ is the normalization constant and $V( {r})$ is a
dimensionless function in the form
\begin{equation}
V( {r})=\frac{1}{\hbar c}\int dr v(r),
\end{equation}
when units are recovered. By assuming~\cite{Jackiw2}
\begin{equation}
v (r)_{\ \overrightarrow{r \to \infty}}\ M,
\end{equation}
where $M$ is a positive definite constant with the units of energy
(when units are recovered, otherwise its unit is the inverse of
length in the natural units) that represents the magnitude of the
pairing potential $|v(r)|$ with the typical value~\cite{Fernandes}
of the order $meV$, one gets
\begin{equation}
\psi^v_0 \approx N\, e^{-i\pi/4}\ e^{-M r}
\left(
\begin{array}{c}
1\\
0
\end{array}
\right).
\label{eq:27aa}
\end{equation}
To the same approximation, the anti-vortex at a fixed distance, say ${\bf r} = {\bf
R}$, leads to
\begin{equation}
\psi^{\bar{v}}_0 \approx N\, e^{i\pi/4}\ e^{-M |{\bf r} - {\bf
R}|} \left(
\begin{array}{c}
0\\
1
\end{array}
\right),
\end{equation}
\end{subequations}
Also, the corresponding 4-spinors that solve Eq.
\eqref{schrodinger} at zero-energy are
\begin{eqnarray}
\Psi^v_0 \approx
\left(
\begin{array}{c}
 N\, e^{-i \pi/4}\, e^{-M {\bf r}}\\[.5ex]
 0 \\[.5ex]
 0\\[.5ex]
 -N\,  e^{i \pi/4}\, e^{-M {\bf r}} 
\end{array}
\right)\qquad\qquad {\rm and}\qquad\qquad \Psi^{\bar{v}}_0 \approx
\left(
\begin{array}{c}
0\\[.5ex]
 N\, e^{i \pi/4}\, e^{-M |{\bf r} - {\bf R}|}\\[.5ex]
 N \,e^{-i \pi/4}\,  e^{-M |{\bf r} - {\bf R}|}  \\[.5ex]
 0\\[.5ex]
\end{array}
\right).
\label{eq:28b}
\end{eqnarray}

It is known that a non-local
complex fermion can be written in terms of two
MFs~\cite{Alicia,Lijense} such that
\begin{equation}\label{eq:kitaevMFannihilation}
    c= \frac{1}{2} \left( \Gamma_{1} + i \Gamma_{2}
    \right)\qquad\quad
    {\rm and}\qquad\quad
    c^\dagger = \frac{1}{2} \left( \Gamma_{1} - i \Gamma_{2} \right),
\end{equation}
where $c$ is the electron annihilation operator and $\Gamma_i$s
are the Majorana operators. This point can be seen more clearly by
inverting relation~(\ref{eq:kitaevMFannihilation}) as
\begin{equation}\label{eq:kitaevMFinv1}
    \Gamma_{1} = c^\dagger + c\qquad\quad {\rm and}\qquad\quad
    \Gamma_{2} = i \left( c^\dagger - c \right),
\end{equation}
which are obviously hermitian operators and hence, the Majorana
operators. Hence, in our system (including a vortex and an
anti-vortex located at a distance far away from each other),
there should also exist two fermionic bound states. One of those
with positive energy and the other one with equal magnitude but
with opposite sign.

Calculation shows that the current density
$j_{\mu}=\bar\Psi_0\gamma_{\mu}\Psi_0$ vanishes for each MZM state
alone. On the other hand, the complex fermionic state composed of
two non-local MZMs, bound to a vortex localized at ${\bf r} = {\bf
R/2}$ and an anti-vortex at ${\bf r} = -{\bf R/2}$, can be written
as
\begin{eqnarray}
\Psi^{v \bar{v} } ({\bf r}) =\Psi^{v }_0({\bf r})+i\, \Psi^{ \bar{v}
}_0({\bf r})=
 N e^{-M |{\bf r} -\frac{{\bf R}}{2}|}\left(
\begin{array}{c}
\, e^{-i(\frac{ \pi}{4}+\frac{\alpha}{2})} \, \\[.5ex]
 0\\
 0\\
 -\,  e^{i( \frac{\pi}{4}+\frac{\alpha}{2})} \,
\end{array}
\right)+
N e^{-M |{\bf r} +\frac{{\bf R}}{2}|}\left(
\begin{array}{c}
 0\\
 i\, e^{i(\frac{ 3\pi }{4}+\frac{\alpha}{2})}    \\
i\, e^{-i(\frac{ 3\pi }{4}+\frac{\alpha}{2})}  \\
 0\\
\end{array}
\right) \label{eq:28bb}
\end{eqnarray}
with the energy $\varepsilon \approx  e^{- M R}$, where $\alpha $
is a constant phase originating from the mutual effect of
vortex/anti-vortex on each other~\cite{Jackiw2,FuKane}. It should
be emphasized that $\varepsilon$ is the energy of the complex
fermionic state composed of the two MZMs and not the energy of
each MZM. Using MZMs properties, i.e., $(\Psi^{{v }}_0)^c({\bf
r})=\Psi^{{v  }}_0({\bf r})$ and $(\Psi^{{ \bar{v} }}_0)^c({\bf
r})=\Psi^{{\bar{v} }}_0({\bf r})$, the complex anti-fermionic
state obviously is
\begin{eqnarray}
(\Psi^{v \bar{{v}}})^c({\bf r}) =\Psi^{{v  }}_0({\bf r})-i\,\Psi^{{
\bar{v} }}_0({\bf r})=
 N e^{-M |{\bf r} -\frac{{\bf R}}{2}|}\left(
\begin{array}{c}
\,  e^{-i(\frac{ \pi}{4}+\frac{\alpha}{2})}\, \\[.5ex]
 0\\
 0\\
 -\,  e^{i( \frac{\pi}{4}+\frac{\alpha}{2})} \,
\end{array}
\right)-
N e^{-M |{\bf r} +\frac{{\bf R}}{2}|}\left(
\begin{array}{c}
 0\\
 i\, e^{i(\frac{ 3\pi }{4}+\frac{\alpha}{2})}    \\
i\, e^{-i(\frac{ 3\pi }{4}+\frac{\alpha}{2})}  \\
 0\\
\end{array}
\right), \label{eq:28bbb}
\end{eqnarray}
which belongs to the energy $\varepsilon \approx - e^{- M R}$.

A wave-packet consisted of both the fermionic fields can be
written as~\cite{Jackiw2}
\begin{equation}
\Phi({\bf r},t)  \equiv \,  \Psi^{v \bar{v}}({\bf r})
e^{-i\varepsilon t} +\,(\Psi^{{v \bar{v}}})^c({\bf r})
e^{i\varepsilon t}, \label{eq:290}
\end{equation}
in which $ \Psi^{v \bar{v} }({\bf r}) $ and $ (\Psi^{{v
\bar{v}}})^c({\bf r})$ respectively represent the positive and
negative energy complex fermionic states. Employing
wave-packet~(\ref{eq:290}) to calculate
$j_{\mu}=q\bar\Phi\gamma_{\mu}\Phi$, with $q$ as the electric
charge, leads to a time-dependent current density, which
originates from the interference of the positive and negative
energy parts. For simplicity, we consider ${\bf R}$ along the
x-axis. Then, we calculate the zero component of this current
density for areas far away from those two MBSs, i.e. when $r\gg
R$, in two limits: for regions close to the axis that joins those
(i.e., for when $\phi\rightarrow 0$) and regions close to the
direction perpendicular to the axis of their connection (i.e., for
when $\phi\rightarrow \pi/2$). The results represent a transient
emergent electric charge described as (when units are recovered)
\begin{eqnarray}\label{chiralmzm1}
j^{ \phi\rightarrow 0}_0= 2q\, e^{-Mr/(\hbar c)} \tanh{(M
R/\hbar c)}\,\cos{(2\varepsilon t)},
\end{eqnarray}
and for regions of $\phi\rightarrow
\pi/2$, it becomes
\begin{eqnarray}\label{chiralmzm2}
j^{ \phi\rightarrow \pi/2 }_0=0.
\end{eqnarray}
The interference term exhibits an oscillatory behavior along the
axis joining the two MBSs, which vanishes only after averaging it
in time. In other words, although the charge of a single MZM is
exactly zero, when two MZMs take part to build a charged fermion,
an oscillatory transient charge (and not a net charge) appears in
the system. This result can give some insight to the
vortex/anti-vortex system and is different from the results of Ref.~\cite{Jackiw2}, wherein emergence of a field of non-vanishing energy was mentioned without specifying the charge of such a field. The amplitude of oscillations fade away
when $R\rightarrow 0$, as expected since this limit corresponds to
the fusion of the vortex and the anti-vortex. The frequency of the
oscillation is proportional to $2e^{- M R}$, which goes to zero in
the limit $R\rightarrow \infty$. As it is shown in Fig.~$1$, due
to the proportionality of the frequency and amplitude of the
oscillations to distances $r$ and $R$, a small change in these
distances makes those change substantially. More specifically, the
amplitude is very sensitive to changes in $r$, $R$ and $M$. Hence
\begin{figure}[ht]\label{fig1}
\includegraphics[width=10cm, height=5cm]{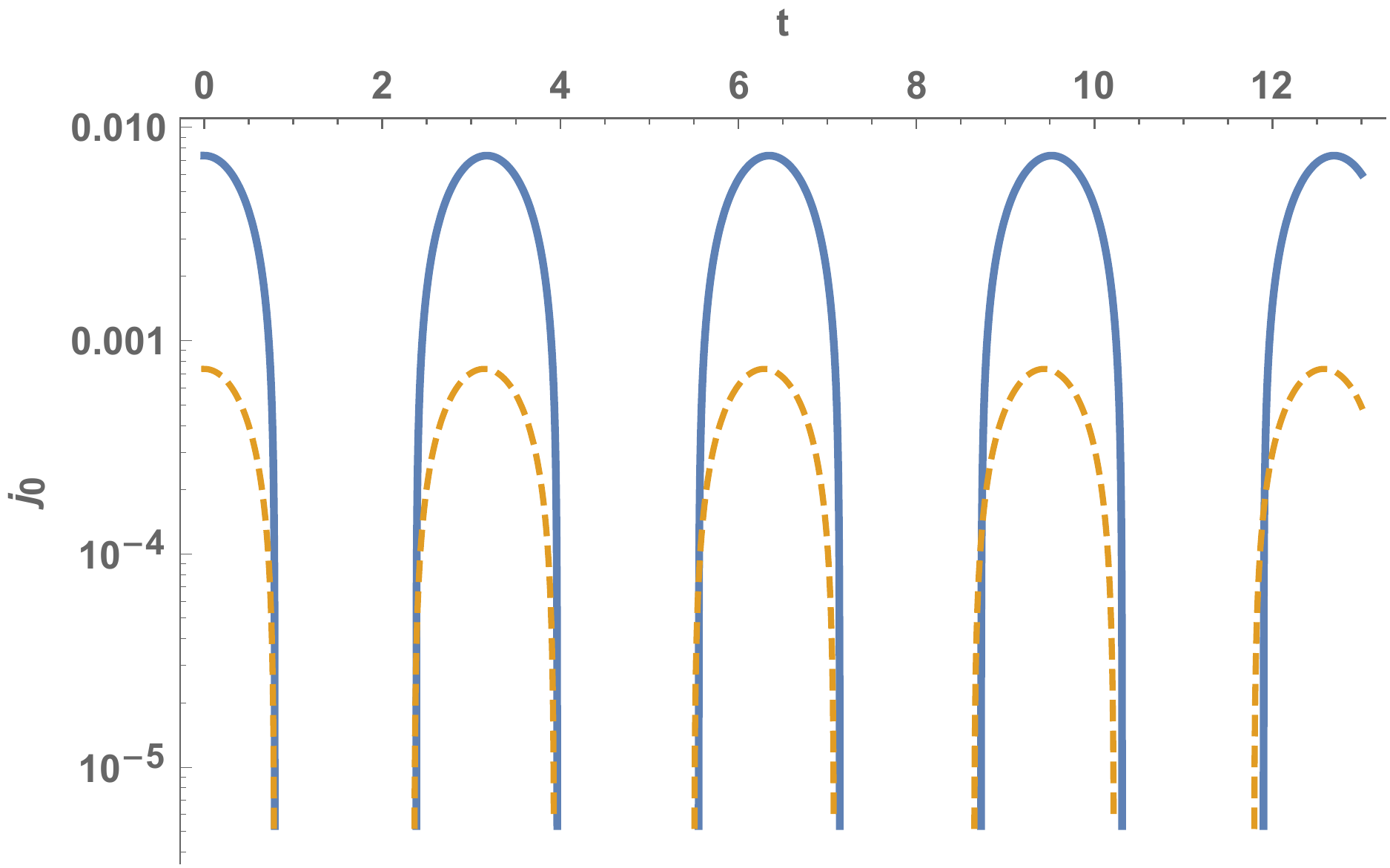}
\centering \caption{\small The chart shows the $j^{\phi\rightarrow
0}_0$ in units of electric charge q versus time in units of
second. The oscillatory patterns are depicted for $M\approx1 meV$ in normal superconductors, and $r\approx10^{-4}m$ such that $M r/(\hbar c)\approx1$, where
the solid line represents $R\approx10^{-6}m$ (i.e. $M
R/(\hbar c)\approx 10^{-2}$) and the dashed line for
$R\approx10^{-7}m$ (i.e. $M R/(\hbar c)\approx 10^{-3}$).}
\end{figure}
for it to be detectable. 
 we consider, say, $ M r/(\hbar
c)\approx1$ or less and subsequently $ M R/(\hbar c)\ll 1$ due
to the $r\gg R$ condition. The values mentioned in the caption would lead to a
detectable setup. 
On the other hand, relation~(\ref{chiralmzm2})
shows that, in the direction perpendicular to the axis connecting
the two MBSs, the current density vanishes identically.

As yet another feature of MBSs, using the interference of the
positive and negative energy components of
wave-packet~(\ref{eq:290}) and after making some calculations, the
spatial current density components, for both $\phi\rightarrow 0$
and $\phi\rightarrow \pi/2$, are
\begin{eqnarray}\label{ZBWvelocity1}
j_1\!\!\!&=&\!\!\! 2q\, e^{-2M r} \cos\alpha\,\sech{(M R)}\,\sin(2
\varepsilon t),\nonumber\\ [.5ex]
 j_2\!\!\!&=&\!\!\! 2q\, e^{-2M r}\sin{\alpha}\,\sech{(M R)}\,\sin(2 \varepsilon
 t).
\end{eqnarray}
These results describe an oscillatory behavior in the $x-y$ plane
and, in principle, should also be detectable for values of the
energy $\varepsilon$ in the range of experimental resolution.

As stated above, the two MBSs lead to a non-local complex fermion
along with its anti-particle. Accordingly, using
wave-packet~(\ref{eq:290}), one can calculate the expectation
value $<\!\alpha\!>$, which corresponds to the ZBW velocity of a
complex fermion when both positive and negative energies are
involved. The result of such a calculation coincides with the
spatial component of the current density given in
relations~(\ref{ZBWvelocity1}). In the other word,
relations~(\ref{ZBWvelocity1}) describe the ZBW motion of the
complex fermion built out of the two MBSs. Also, as the involved
complex fermion is non-local in nature, the FW transformation
would~not be relevant to this case. Besides, one of the main
obstacles regarding the detection of the ZBW of electron (i.e.,
the ultrahigh frequency of the oscillations) would be absent in
the MBSs setup because the frequency can be adjusted. However it
should be mentioned, that if a system of two MZMs bound to two
vortices instead of a vortex and an anti-vortex, then no
oscillatory pattern will emerge in the system. It is interesting
to note that these results might be applicable to the one
dimensional models as well~\cite{Kitaev}.

\section{Conclusions}
\indent

The push to fabricate topologically protected qubits using MZMs
(which are predicted to emerge as localized zero-energy bound
states in topological superconductors) is one of the most
appealing research topics in quantum condensed matter physics.
However, to create MZMs in practice it needs the combination of
cutting-edge fields such as the nanotechnology, superconductivity,
the device engineering and materials science. In other words,
being charge-less and its own anti-particle makes it difficult to
take a MZM out of the sample for private inspection. These
problems have led to redesign experiments in such a way that to
probe other features of MZMs in indirect detection, for example
via measuring the energy of MBS systems.

In this work, we have taken advantage of locating two MBSs at a
distance from each other in order to be able properly describe a
non-local complex fermionic state. Accordingly, we have computed
the current density of a wave-packet consisted of the complex
fermionic state along with its anti-particle with opposite energy. 
We have managed to specify that although MZMs break the $U(1)$ gauge symmetry in general, 
 a non-vanishing conserved
oscillatory transient charge emerges, as shown in Figure~$1$, which vanishes in the
direction perpendicular to the axis joining the two MBSs. The
frequency of this oscillation and its amplitude are proportional
to the distance between the two MBSs. Hence, by adjusting this
distance, such a current density would be detectable. Moreover,
the spatial components of the current density exhibit an
oscillatory behavior with the same property, which is also a
reminiscent of the ZBW effect. In addition, this oscillatory
behavior should also be detectable for values of energy in the
range of experimental resolution.

The ZBW phenomenon is a long sought effect which has eluded the
experimental observation so far due to its extremely high
frequency. In this regard, it has even been doubted to be physical
due to the FW transformation, which eliminates negative energy
components in electron wave functions. In this work, we have
demonstrated that the emerged ZBW motion originates from the
interference of the positive and negative energy parts of the
fermionic states constructed from the two non-local MBSs
(correlations incompatible with a local hidden variable theory).
Therefore,  the FW transformation would not be relevant to this
case. Furthermore, those measurements that indicate the existence
of the current oscillations of the MBSs in the vortex and
anti-vortex background, on the surface of topological insulators
in contact with an s-wave superconductor, would be a verdict
confirmation of the existence of the MZMs and the ZBW phenomenon
at the same time and point to an existing direction to help
unravel the mystery of ZBW.


%
\end{document}